\documentclass{article}

\usepackage{arxiv}

\usepackage[utf8]{inputenc} 
\usepackage[T1]{fontenc}    
\usepackage{hyperref}       
\usepackage{url}            
\usepackage{booktabs}       
\usepackage{amsfonts}       
\usepackage{nicefrac}       
\usepackage{microtype}      
\usepackage{lipsum}
\usepackage{amssymb,amsmath,bm}
\usepackage[dvips,dvipdfmx]{graphicx}
\usepackage{color,xcolor}

\usepackage{tabularx}
\usepackage{multirow}
\definecolor{blue}{rgb}{ 0    0.3470    0.7410}
\definecolor{red}{rgb}{0.8500, 0.1250, 0.0480} 
\definecolor{orange}{rgb}{0.8500, 0.3250, 0.0980} 
\definecolor{yellow}{rgb}{0.9290, 0.6940, 0.1250}
\definecolor{purple}{rgb}{0.4940, 0.1840, 0.5560}
\definecolor{gray}{rgb}{0.52, 0.52, 0.6}
\definecolor{green}{rgb}{0.0, 0.4, 0.0}

\newcommand{\bd}[1]{\mbox{\boldmath$#1$}}

\newcommand{\Pd}[1]{\ensuremath{\Bigl[\!\!\Bigl[#1\Bigr]\!\!\Bigr]}}

\title{Adjoint-based phase reduction analysis of incompressible periodic flows}
\author{
Yoji Kawamura \\
  Center for Mathematical Science and Advanced Technology\\
  Japan Agency for Marine-Earth Science and Technology\\
 Yokohama 236-0001, Japan \\
  \texttt{ykawamura@jamstec.go.jp} \\
   \And
 Vedasri Godavarthi \\
Department of Mechanical and Aerospace Engineering\\
University of California, Los Angeles\\
CA 90095, USA \\
  \texttt{vedasrig@g.ucla.edu} \\
  \And
Kunihiko Taira \\
Department of Mechanical and Aerospace Engineering\\
University of California, Los Angeles\\
CA 90095, USA \\
  \texttt{ktaira@seas.ucla.edu} \\
}

\begin{document}
\maketitle
\begin{abstract}
We establish the theoretical framework for adjoint-based phase reduction analysis for incompressible periodic flows.  Through this adjoint-based method, we obtain spatiotemporal phase sensitivity fields through a single pair of forward and backward direct numerical simulations, as opposed to the impulse-based method that requires a very large number of simulations.  Phase-based analysis involves perturbation analysis about a periodically varying base state and hence is tailored for the analysis of periodic flows.  We formulate the phase description of periodic flows with respect to the potential and vortical perturbations in the flow field.  The current phase-reduction analysis can also be implemented consistently in the immersed boundary projection method, which facilitates the analysis over arbitrarily-shaped bodies.  We demonstrate the strength of the phase-based analysis for periodic flows over circular cylinder and symmetric airfoils at high incidence angles.  The critical regions for phase modification in the cylinder flow are investigated and the locations of flow separation are shown to be the most sensitive regions.  Further, the results reveal the influence of the angle of attack and airfoil thickness on the phase-sensitivity distribution of flows over various airfoils.  The phase for such flows is defined based on the lift coefficient, and hence is influenced by the vortical structures responsible for lift production.  The present framework sheds light on the connection between phase-sensitivity and vortex formation dynamics.  
\end{abstract}


\section{Introduction} \label{sec:1}

In an effort to understand the dynamical richness and complexity of fluid flows, analysis of flow unsteadiness has been a major focus of fluid mechanics research.  The predominant focus of tackling unsteady fluid flows has been understanding the instability mechanisms woven in the overall flow features, producing a wealth of knowledge on various flow instabilities and transition processes.  A flow instability is characterized by a growth of perturbation over a base state.  Local and global stability analyses of a range of flows have been performed about some time-invariant base states~\cite{ref:theofilis11,ref:schmid01}.  With the assumption of small perturbations about these states, the linearized Navier-Stokes equations can be cast in the form of eigenvalue problems and the flow instability is determined by their spectral characteristics.  Such theoretical and numerical techniques developed over the past few decades have enabled us to find instability mechanisms to identify the emergence of linear instabilities like Kelvin-Helmholtz instability, transition to turbulence and development of optimal control techniques based on the growth of perturbations~\cite{ref:drazin04,ref:schmid01,ref:theofilis11,ref:luhar14,ref:cayeh19}.  Such endeavors have more recently been reinforced with data-driven techniques~\cite{ref:holmes12,ref:schmid10,ref:tairamodal17,ref:taira20,ref:hermann21}.  

For the aforementioned studies, the base states have generally been time-invariant.  However, there is a large class of fluid flow problems for that has a high level of unsteadiness in the base flow.  For instance, unsteady wake dynamics of flow over bodies, such as flows over a circular cylinder or airfoils, are characterized by periodic vortex shedding.  Such flows are characterized by time-periodic base flow upon which secondary unsteadiness can grow or decay.  For such a periodic base flow, Floquet analysis~\cite{ref:herbert87} serves essentially as the sole work horse to analyze the behavior of perturbations.  In these analyses, the key concern of the methods have been on the growth in the amplitude of the perturbations and the control strategies aimed at reduction in the amplitude of these oscillations.  There has, in contrast, been limited discussions on the effect of timing on the evolution of these perturbations.  

\begin{figure}
    \centering
    \includegraphics[width=0.8\textwidth]{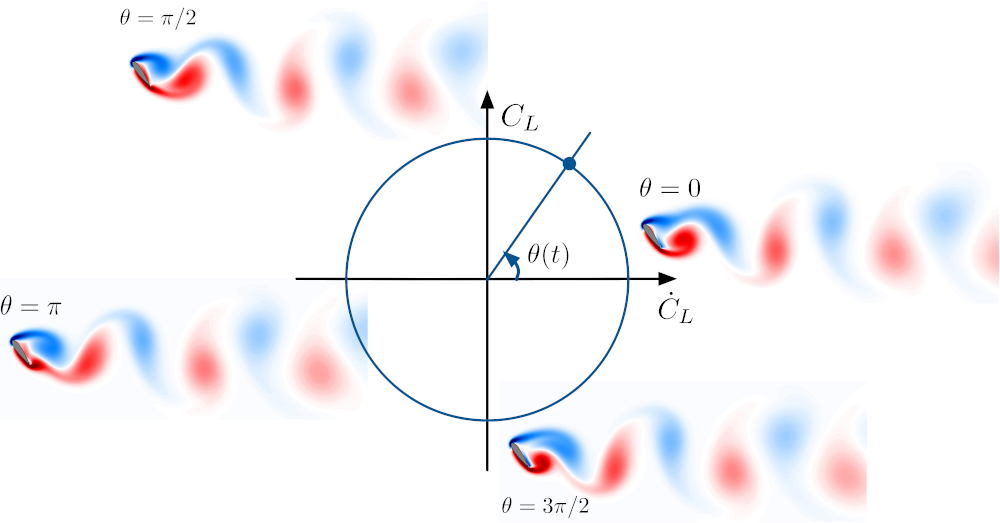}
    \caption{Phase $\theta(t)$ defined over the $C_L-\dot{C}_L$ plane for flow over a NACA0020 airfoil at $\alpha = 55^\circ$ and $Re = 100$.}
    \label{fig:phase_definition}
\end{figure}

For periodic systems, the temporal dynamics can be associated with a phase defined based on the limit-cycle oscillations.  Such phase-based analysis techniques for dynamics about time-periodic systems can play a role in revealing the sensitivity of the periodically varying base flow to certain type of added perturbations.  Essentially, we are then concerned with the evolution of phase, a single scalar variable of the system instead of high-dimensional dynamics which significantly simplifies the complexity associated with unsteady flow analysis.  For instance, the temporal evolution of periodic vortex shedding of flow over an airfoil can be represent as phase evolution using the limit-cycle oscillations of the lift coefficient as shown in figure~\ref{fig:phase_definition}.  We anticipate that this kind of phase-based analysis will become increasingly important as our interests in the analysis, modeling, estimation, and control of unsteady aerodynamics continue to grow.  

Phase reduction analysis has been successfully applied to the analysis of dynamics and synchronization phenomena among the limit-cycle oscillators for various biological, chemical, and neural systems~\cite{ref:winfree80,ref:kuramoto84,ref:pikovsky01,ref:hoppensteadt97,ref:izhikevich07,ref:ermentrout10,ref:ermentrout96,ref:brown04,ref:nakao16,ref:pietras19}.  Through phase reduction, the time evolution of a system is described by a phase equation, which not only facilitates theoretical analysis but also allows for fundamental understanding~\cite{ref:winfree80,ref:kuramoto84,ref:pikovsky01,ref:strogatz00,ref:acebron05,ref:arenas08,ref:pikovsky15,ref:rodrigues16,ref:ashwin16,ref:stankovski17}.  Extracting the phase dynamics can reveal the synchronization characteristics and designing optimal control can modify the phase dynamics to lock-in and synchronize to external frequency.  In addition, phase reduction analysis is also sensor friendly as it only requires temporal measurements capturing the phase of the limit-cycle oscillation.  Hence phase-based analysis is remarkably useful for analysis and control of periodic fluid flows, however such applications have been very recent~\cite{ref:kawamura13,ref:kawamura14,ref:kawamura15,ref:kawamura19,ref:taira18,ref:khodkar20,ref:khodkar21,ref:nair21,ref:loe21,ref:skene22,ref:iima19,ref:iima21}.  

The phase dynamics of a system obtained from the phase-reduction analysis is characterized by the phase sensitivity function, which quantifies the phase response of the system to perturbations.  Spatial phase sensitivity functions extracted from periodic fluid flows identify critical regions in the flow field that can cause significant modification in phase dynamics when perturbed.  Hence these sensitivity fields can be used to design optimal forcing for control of phase dynamics and thereby the flow field characteristics.  There are mainly two methods to obtain the phase sensitivity function~\cite{ref:winfree80,ref:kuramoto84,ref:pikovsky01,ref:hoppensteadt97,ref:izhikevich07,ref:ermentrout10,ref:ermentrout96,ref:brown04,ref:nakao16,ref:pietras19}: 
one is the direct method using impulsive perturbation applied to the system, 
and the other is the adjoint method using the adjoint equations derived from the governing equations of the system.  
The direct method has been used to characterize and control the periodic vortex shedding around cylinders~\cite{ref:taira18,ref:khodkar20,ref:khodkar21,ref:nair21,ref:loe21} and airfoil~\cite{ref:nair21}.  The lock-on characteristics of vortex shedding for a circular cylinder to various periodic perturbations such as, periodic external forcing~\cite{ref:taira18,ref:khodkar20}, periodic vibrations of cylinder~\cite{ref:khodkar21} and fluid-structure interactions~\cite{ref:loe21} are demonstrated via phase reduction using the direct method.  Subsequently, Nair et al.~\cite{ref:nair21} developed a transient control technique based on the phase sensitivity function obtained via the direct method to modify the phase of vortex shedding behavior of flows past a circular cylinder and an airfoil.  Thus, phase sensitivity fields reveal essential physics responsible for optimal control design for periodic fluid flows.  However, the direct method when used to obtain phase sensitivity function, involves obtaining phase response by perturbing the flow with impulse perturbations at several locations in the flowfield, and at different phases of the time period.  Therefore, obtaining spatial phase sensitivity fields with respect to various kinds of perturbations is computationally expensive.  

Alternatively, the adjoint-based method involves solving the adjoint equation derived from the governing equation of the sy2stem.  Thereby, adjoint-based method results in spatial phase sensitivity fields corresponding to perturbations with respect to various state variables through a single pair of the forward computation of governing equations resulting in limit-cycle oscillations and the backward computation of adjoint equation for the corresponding time period. Adjoint method has seen a few applications in periodic flows but limited to Hele-Shaw convection~\cite{ref:kawamura13,ref:kawamura14,ref:kawamura15}, Rayleigh-Bernard convection~\cite{ref:kawamura19} and thermoacoustic oscillations~\cite{ref:skene22}.  However, most of these applications are based on the adjoint formulation of reduced-order governing equations of incompressible flows.  There is a need for a rigorous generalizable theoretic and computational framework for the adjoint-based formulation for the complete incompressible Navier-Stokes equations.  

In this work, we provide a rigorous theoretic framework for adjoint-based phase reduction for incompressible periodic flows.  The evolution of phase sensitivity function is derived, which can be solved with any well-established numerical scheme.  Further, the properties of the phase sensitivity function and the relationship of phase sensitivity field corresponding to perturbations with respect to velocity potential, velocity, and vorticity are presented. In particular, we demonstrate the adjoint-based phase sensitivity for vortex shedding behind canonical bodies.  We show that the adjoint framework can be applied consistently using the immersed boundary projection method~\cite{ref:taira07} which can simulate flows over stationary or moving bodies with arbitrary shapes. We demonstrate this analysis for von Karman vortex street over a circular cylinder and symmetric airfoils at high angles of attack. This formulation enables us to identify the critical regions in the flow field that facilitate the modification of vortex formation process. We then discuss the open loop control strategies that result in lift enhancement by modifying the vortex formation dynamics. This work paves way to control the flow physics in a computationally efficient manner.  

The present paper is organized as follows. The theoretical framework to develop the adjoint-based phase description of periodic fluid flows using incompressible Navier-Stokes equation is described in Sec.~\ref{sec:2}. The demonstration of this adjoint-based framework to analyze the phase sensitivity fields of flows over a circular cylinder and symmetric airfoils at high angles of incidence simulated using the immersed boundary projection approach is presented in Sec.~\ref{sec:3}. In Sec.~\ref{sec:4}, we provide concluding remarks on the present work and possible extensions.

\section{Phase description of periodic flows} \label{sec:2}

Let us present the theoretical framework for the adjoint-based phase description of periodic flows. This section presents the limit-cycle solutions and the derivation of phase sensitivity functions with respect to velocity-based, vector potential-based, and vorticity-based perturbations.  We comment on the properties of the phase sensitivity functions and the relation among the different phase sensitivity fields. While the focus of this paper is placed on incompressible flows, the approach herein can be applied to compressible flows without any difficulty. To develop the framework for adjoint-based phase description for periodic flows, we first consider the limit-cycle solution to the incompressible Navier-Stokes equations and we linearize the dynamics for a periodic perturbation about the considered limit-cycle solution.  

\subsection{Adjoint analysis of periodic incompressible flows}

The non-dimensional incompressible Navier-Stokes equations can be expressed as
\begin{align}
  \hat{M} \frac{\partial}{\partial t} \bd{q}(\bd{x}, t)
  = \bd{\cal F}[\bd{q}],
  \label{eq:q}
\end{align}
where
\[
  \bd{q}(\bd{x}, t) = 
  \left(\begin{array}{c}
    \bd{u} \\
    p
  \end{array}\right),
  \qquad
  \hat{M} = 
  \begin{bmatrix}
    1&&& \\
    &1&& \\
    &&1& \\
    &&&0
  \end{bmatrix},
  \qquad
  \bd{\cal F}[\bd{q}] = 
  \begin{cases}
    \displaystyle
    - \bd{u} \cdot \nabla \bd{u}
    - \nabla p
    + {\rm Re}^{-1} \nabla^2 \bd{u},
    \\[3mm]
    \displaystyle
    \nabla \cdot \bd{u}.
  \end{cases}
\]
Here, $\bd{u}$ and $p$ are the velocity and pressure fields and $Re$ is the Reynolds number.
In general,
a stable limit-cycle solution of Eq.~(\ref{eq:q}),
which represents a periodic flow $\tilde{\bd{q}}$,
can be described as
\begin{align}
  \bd{q}(\bd{x}, t) = \tilde{\bd{q}}(\bd{x}, \theta(t)),
  \qquad
  \dot{\theta}(t) = \omega_n,
\end{align}
where $\theta$ and $\omega_n$ are the phase and frequency of the periodic flow, respectively.
The phase $\theta$ describes the periodic behavior of the system.
The evolution of phase along the limit cycle for a periodic flow can be defined based on state-space formed with temporal measurements that characterize the flow physics.
For instance, $C_L-\dot{C}_L$ plane can be used to capture the periodic vortex shedding over NACA0020 airfoil at $\alpha = 55^\circ$ and $Re = 100$ as shown in figure~\ref{fig:phase_definition}.
(While $\theta$ is defined along the limit cycle, we can define $\Theta(\bd{q})$ as the phases of the state variables $\bd{q}$ in the vicinity of the limit cycle. Since $\omega_n$ is the natural frequency of the system, $\dot{\Theta}(\bd{q}) = \nabla \Theta(\bd{q}) \cdot \dot{\bd{q}} = \omega_n$ and we can analyze the perturbation dynamics in the neighborhood of the limit cycle using the evolution of $\Theta(\bd{q})$~\cite{ref:taira18}.)
Substituting $\tilde{\bd{q}}$ into Eq.~(\ref{eq:q}),
we find that $\tilde{\bd{q}}(\bd{x}, \theta)$ satisfies the following equation
\begin{align}
  \omega_n \hat{M} \frac{\partial}{\partial \theta} \tilde{\bd{q}}(\bd{x}, \theta)
  = \bd{\cal F}[\tilde{\bd{q}}].
  \label{eq:q-q0}
\end{align}
We now introduce a small disturbance $\bd{q}^\prime(\bd{x}, \theta, t)$ to $\tilde{\bd{q}}(\bd{x}, \theta)$ as
\begin{align}
  \bd{q}(\bd{x}, t) = \tilde{\bd{q}}(\bd{x}, \theta) + \bd{q}^\prime(\bd{x}, \theta, t).
  \label{eq:q-prime}
\end{align}
Equation~(\ref{eq:q}) is then linearized with respect to $\bd{q}^\prime(\bd{x}, \theta, t)$ as follows:
\begin{align}
  \hat{M} \frac{\partial}{\partial t} \bd{q}^\prime(\bd{x}, \theta, t)
  = \hat{\cal L}(\bd{x}, \theta) \bd{q}^\prime(\bd{x}, \theta, t),
  \label{eq:q-q1}
\end{align}
where the linear operator $\hat{\cal L}(\bd{x}, \theta)$ and the eigenfunction $\bd{q}^\prime(\bd{x}, \theta)$ is 
\begin{align}
  \hat{\cal L}(\bd{x}, \theta) \bd{q}^\prime(\bd{x}, \theta)
  = \left[ \hat{\cal J}(\bd{x}, \theta) - \omega_n \hat{M} \frac{\partial}{\partial \theta} \right]
  \bd{q}^\prime(\bd{x}, \theta),
  \label{eq:calL}
\end{align}
and the components of $\hat{\cal J}(\bd{x}, \theta) \bd{q}^\prime(\bd{x}, \theta)$ are
\begin{align}
  \hat{\cal J} \bd{q}^\prime = 
  \begin{cases}
    \displaystyle
    - \bd{u}^\prime \cdot \nabla \tilde{\bd{u}}
    - \tilde{\bd{u}} \cdot \nabla \bd{u}^\prime
    - \nabla p^\prime
    + {\rm Re}^{-1} \nabla^2 \bd{u}^\prime,
    \\[3mm]
    \displaystyle
    \nabla \cdot \bd{u}^\prime.
  \end{cases}
  \label{eq:calJ}
\end{align}
In Eq.~(\ref{eq:calL}), we omitted the $t$-dependence of $\bd{q}^\prime(\bd{x}, \theta, t)$ and denoted it as $\bd{q}^\prime(\bd{x}, \theta)$, because we consider only the eigenvalue problem of the linear operator $\hat{\cal L}(\bd{x}, \theta)$, and therefore, the $t$-dependence of $\bd{q}^\prime$ do not appear hereafter. We note that not only the limit-cycle solution $\tilde{\bd{q}}(\bd{x}, \theta)$ but also the eigenfunction $\bd{q}^\prime(\bd{x}, \theta)$ satisfy the $2\pi$-periodicity with respect to $\theta$:
\begin{align}
  \tilde{\bd{q}}(\bd{x}, \theta + 2\pi) = \tilde{\bd{q}}(\bd{x}, \theta),
  \qquad
  \bd{q}^\prime(\bd{x}, \theta + 2\pi) = \bd{q}^\prime(\bd{x}, \theta).
  \label{eq:periodicity}
\end{align}

We now introduce the adjoint variables required to derive the phase equation and phase sensitivity function.
We define the adjoint of the perturbation as ${\bd{q}^\prime}^\ast(\bd{x}, \theta)$.
Similar to $\bd{q}^\prime$, ${\bd{q}^\prime}^\ast$ also satisfies $2\pi$-periodicity with respect to $\theta$ as ${\bd{q}^\prime}^\ast(\bd{x}, \theta + 2\pi) = {\bd{q}^\prime}^\ast(\bd{x}, \theta)$.
The inner product of the adjoint with itself is defined in two ways.
First, we define the inner product of two functions over space as
\begin{align}
  \Bigl\langle {\bd{q}^\prime}^\ast(\bd{x}, \theta), \, \bd{q}^\prime(\bd{x}, \theta) \Bigr\rangle
  \equiv \int_D \,
  \Bigl[ {\bd{u}^\prime}^\ast(\bd{x}, \theta) \cdot \bd{u}^\prime(\bd{x}, \theta) + {p^\prime}^\ast(\bd{x}, \theta) p^\prime(\bd{x}, \theta) \Bigr] \,
  d\bd{x}.
  \label{eq:inner-product}
\end{align}
Second,
using Eq.~(\ref{eq:inner-product}),
we also define the inner product of two functions as
\begin{align}
  \Pd{ {\bd{q}^\prime}^\ast(\bd{x}, \theta), \, {\bd{q}^\prime}(\bd{x}, \theta) }
  \equiv \frac{1}{2\pi} \int_0^{2\pi} \,
  \Bigl\langle {\bd{q}^\prime}^\ast(\bd{x}, \theta), \, \bd{q}^\prime(\bd{x}, \theta) \Bigr\rangle \, d\theta.
  \label{eq:Inner-Product}
\end{align}

Using Eq.~(\ref{eq:Inner-Product}),
we introduce the adjoint operator of $\hat{\cal L}(\bd{x}, \theta)$ as
\begin{align}
  \Pd{ {\bd{q}^\prime}^\ast(\bd{x}, \theta), \, \hat{\cal L}(\bd{x}, \theta) {\bd{q}^\prime}(\bd{x}, \theta) } = 
  \Pd{ \hat{\cal L}^\ast(\bd{x}, \theta) {\bd{q}^\prime}^\ast(\bd{x}, \theta), \, \bd{q}^\prime(\bd{x}, \theta) } + 
  {\cal S}\left[ {\bd{q}^\prime}^\ast(\bd{x}, \theta), \, \bd{q}^\prime(\bd{x}, \theta) \right].
  \label{eq:adjoint-operator}
\end{align}
Here, the bilinear concomitant is denoted by
${\cal S}[ {\bd{q}^\prime}^\ast(\bd{x}, \theta), \, \bd{q}^\prime(\bd{x}, \theta) ]$.
Using partial integration,
the adjoint operator $\hat{\cal L}^\ast(\bd{x}, \theta)$ can be expressed as
\begin{align}
  \hat{\cal L}^\ast(\bd{x}, \theta) {\bd{q}^\prime}^\ast(\bd{x}, \theta)
  = \left[ \hat{\cal J}^\ast(\bd{x}, \theta) + \omega_n \hat{M} \frac{\partial}{\partial \theta} \right]
  {\bd{q}^\prime}^\ast(\bd{x}, \theta).
  \label{eq:calLast}
\end{align}
Here, the components of $\hat{\cal J}^\ast(\bd{x}, \theta) {\bd{q}^\prime}^\ast(\bd{x}, \theta)$ are 
\begin{align}
  \hat{\cal J}^\ast {\bd{q}^\prime}^\ast = 
  \begin{cases}
    \displaystyle
    - {u_{x}^\prime}^\ast\nabla \tilde{u}_x 
    - {u_{y}^\prime}^\ast\nabla \tilde{u}_y
    - {u_{z}^\prime}^\ast\nabla \tilde{u}_z
    + \tilde{\bd{u}} \cdot \nabla {\bd{u}^\prime}^\ast
    - \nabla {p^\prime}^\ast
    + {\rm Re}^{-1} \nabla^2 {\bd{u}^\prime}^\ast,
    \\[3mm]
    \displaystyle
    \nabla \cdot {\bd{u}^\prime}^\ast,
  \end{cases}
  \label{eq:calJast}
\end{align}
and the bilinear concomitant
${\cal S}[ {\bd{q}^\prime}^\ast(\bd{x}, \theta), \, \bd{q}^\prime(\bd{x}, \theta) ]$
is given by
\begin{align}
  {\cal S} = 
  - \frac{1}{2\pi} \int_0^{2\pi} \int_{\partial D} \, \bd{\rm n} \cdot \Bigl[
    & \left( {\bd{u}^\prime}^\ast \cdot \bd{u}^\prime \right) \tilde{\bd{u}}
    + {\rm Re}^{-1} \left( u_{x}^\prime \nabla {u^\prime}^\ast_x + u_{y}^\prime \nabla {u^\prime}^\ast_y + u_{z}^\prime \nabla {u^\prime}^\ast_z
    - {u^\prime}^\ast_x \nabla u_{x}^\prime - {u^\prime}^\ast_y \nabla u_{y}^\prime - {u^\prime}^\ast_z \nabla u_{z}^\prime \right)
    \nonumber \\
    &+ p^\prime {\bd{u}^\prime}^\ast - {p^\prime}^\ast \bd{u}^\prime \Bigr] \, dS \, d\theta
  - \frac{1}{2\pi} \int_D \,
  \Bigl[ \omega_n \left( {\bd{u}^\prime}^\ast \cdot \bd{u}^\prime \right) \Bigr]_{\theta = 0}^{ 2\pi} \, d\bd{x},
  \label{eq:calS}
\end{align}
with ${\cal S} = 0$ for the adjoint boundary conditions.
Further details on the boundary conditions are given in Sec.~\ref{sec:boundary}.

\subsection{Zero eigenfunctions and their normalization condition}

We need to consider the Floquet and adjoint eigenfunctions and seek their normalization condition to derive the phase sensitivity fields for periodic fluid flows.
In the calculation that will be performed to obtain the phase equation in Sec.~\ref{sec:phase},
we use the Floquet and adjoint eigenfunctions associated with the zero eigenvalue
of $\hat{\cal L}(\bd{x}, \theta)$ and $\hat{\cal L}^\ast(\bd{x}, \theta)$.
The zero eigenfunctions,
$\tilde{\bd{Q}}(\bd{x}, \theta)$ and $\tilde{\bd{Q}}^\ast(\bd{x}, \theta)$,
satisfy the following conditions:
\begin{align}
  \hat{\cal L}(\bd{x}, \theta) \tilde{\bd{Q}}(\bd{x}, \theta) = 0,
  \qquad
  \hat{\cal L}^\ast(\bd{x}, \theta) \tilde{\bd{Q}}^\ast(\bd{x}, \theta) = 0.
  \label{eq:Q0ast}
\end{align}
The components of the Floquet zero eigenfunction $\tilde{\bd{Q}}(\bd{x}, \theta)$ can be defined as
$(\tilde{\bd{U}}, \, \tilde{P})^{\rm T}$.
The components of the adjoint zero eigenfunction $\tilde{\bd{Q}}^\ast(\bd{x}, \theta)$ can also be defined as
$(\tilde{\bd{U}}^\ast, \, \tilde{P}^\ast)$.
The Floquet zero eigenfunction $\tilde{\bd{Q}}(\bd{x}, \theta)$ can be chosen by differentiating Eq.~(\ref{eq:q-q0}) with respect to $\theta$.
\begin{align}
  \tilde{\bd{Q}}(\bd{x}, \theta)
  = \frac{\partial}{\partial \theta} \tilde{\bd{q}}(\bd{x}, \theta),
  \label{eq:Q0-q0}
\end{align}
which can be confirmed by 
Using Eq.~(\ref{eq:Inner-Product}),
the adjoint zero eigenfunction $\tilde{\bd{Q}}^\ast(\bd{x}, \theta)$ can be normalized as
\begin{align}
  \Pd{ \tilde{\bd{Q}}^\ast(\bd{x}, \theta), \, \hat{M} \tilde{\bd{Q}}(\bd{x}, \theta) }
  = \frac{1}{2\pi} \int_0^{2\pi} \,
  \Bigl\langle \tilde{\bd{Q}}^\ast(\bd{x}, \theta), \, \hat{M} \tilde{\bd{Q}}(\bd{x}, \theta) \Bigr\rangle \, d\theta
  = 1.
  \label{eq:Normalization}
\end{align}
Note that the diagonal matrix $\hat{M}$ has been inserted in Eq.~(\ref{eq:Normalization})
because of the form of Eq.~(\ref{eq:q}).
We also note that the following condition is satisfied as
\begin{align}
  \omega_n \frac{\partial}{\partial \theta}
  \Bigl\langle \tilde{\bd{Q}}^\ast(\bd{x}, \theta), \, \hat{M} \tilde{\bd{Q}}(\bd{x}, \theta) \Bigr\rangle
  &= \Bigl\langle \tilde{\bd{Q}}^\ast(\bd{x}, \theta), \, 
  \omega_n \hat{M} \frac{\partial}{\partial \theta} \tilde{\bd{Q}}(\bd{x}, \theta) \Bigr\rangle
  + \Bigl\langle \omega_n \hat{M}^{\rm T} \frac{\partial}{\partial \theta} \tilde{\bd{Q}}^\ast(\bd{x}, \theta), \, 
  \tilde{\bd{Q}}(\bd{x}, \theta) \Bigr\rangle
  \nonumber \\
  &= \Bigl\langle \tilde{\bd{Q}}^\ast(\bd{x}, \theta), \, 
  \hat{\cal J}(\bd{x}, \theta) \tilde{\bd{Q}}(\bd{x}, \theta) \Bigr\rangle
  - \Bigl\langle \hat{\cal J}^\ast(\bd{x}, \theta) \tilde{\bd{Q}}^\ast(\bd{x}, \theta), \, 
  \tilde{\bd{Q}}(\bd{x}, \theta) \Bigr\rangle
  = 0.
  \label{eq:property}
\end{align}
Therefore,
the following normalization condition is satisfied for every $\theta$:
\begin{align}
  \Bigl\langle \tilde{\bd{Q}}^\ast(\bd{x}, \theta), \, \hat{M} \tilde{\bd{Q}}(\bd{x}, \theta) \Bigr\rangle = 1.
  \label{eq:normalization}
\end{align}
This normalization condition is essential to project the dynamics onto the limit-cycle solution as will be seen in Sec.~\ref{sec:phase}.
For this, we need to compute the adjoint zero eigenfunction $\tilde{\bd{Q}}$ numerically.
According to Eq.~(\ref{eq:Q0ast}), the adjoint zero eigenfunction $\tilde{\bd{Q}}^\ast(\bd{x}, \theta)$ satisfies
\begin{align}
  -\omega_n \hat{M}^{\rm T} \frac{\partial}{\partial \theta} \tilde{\bd{Q}}^\ast(\bd{x}, \theta)
  = \hat{\cal J}^\ast(\bd{x}, \theta) \tilde{\bd{Q}}^\ast(\bd{x}, \theta).
  \label{eq:adjoint-1}
\end{align}
By substituting $\theta = -\omega_n s$,
the above equation can be transformed as
\begin{align}
  \hat{M}^{\rm T} \frac{\partial}{\partial s} \tilde{\bd{Q}}^\ast(\bd{x}, -\omega_n s)
  = \hat{\cal J}^\ast(\bd{x}, -\omega_n s) \tilde{\bd{Q}}^\ast(\bd{x}, -\omega_n s).
  \label{eq:adjoint-2}
\end{align}
Therefore, the adjoint equation can be written in the following form:
\begin{align}
  \frac{\partial}{\partial s} \tilde{\bd{U}}^\ast(\bd{x}, -\omega_n s)
  &=
  - \tilde{{U}}_x^\ast \cdot \nabla \tilde{{u}}_x
  - \tilde{{U}}_y^\ast \cdot \nabla \tilde{{u}}_y
  - \tilde{{U}}_z^\ast \cdot \nabla \tilde{{u}}_z
  + \tilde{\bd{u}} \cdot \nabla \tilde{\bd{U}}^\ast
  - \nabla \tilde{P}^\ast
  + {\rm Re}^{-1} \nabla^2 \tilde{\bd{U}}^\ast,
  \label{eq:U1ast-V1ast} \\
  0
  &= \nabla \cdot \tilde{\bd{U}}^\ast.
  \label{eq:P1ast} 
\end{align}
According to Eq.~(\ref{eq:normalization}),
the normalization condition is provided by
\begin{align}
  \int_D \, \tilde{\bd{U}}^\ast(\bd{x}, \theta) \cdot \tilde{\bd{U}}(\bd{x}, \theta) \, d\bd{x} = 1.
  \label{eq:normalization-U}
\end{align}
Based on Eq.~(\ref{eq:Q0-q0}),
the Floquet zero eigenfunction is given by
\begin{align}
  \tilde{\bd{U}}(\bd{x}, \theta)
  = \frac{\partial}{\partial \theta} \tilde{\bd{u}}(\bd{x}, \theta).
  \label{eq:U1-u0}
\end{align}
From a viewpoint of numerical analysis,
Eq.~(\ref{eq:q}) is similar to a set of Eqs.~(\ref{eq:U1ast-V1ast}) and (\ref{eq:P1ast}).
The differences exist only in the explicit forms of the boundary conditions and advection term.
Therefore, any well-established numerical schemes for the Navier-Stokes equations can also be utilized for its adjoint system.

\subsection{Phase equation and phase sensitivity function of periodic flows} \label{sec:phase}

The phase sensitivity fields encode the sensitivity of periodic flows with respect to perturbations.
Hence, we now consider periodic flows with a weak perturbation $\epsilon \bd{K}(\bd{x}, t)$ added to Eq.~(\ref{eq:q}).
We introduce a variable $\bd{E}(\bd{x}, t) = (\bd{K}, \, 0)^{\rm T}$.
Thus, we can express the Navier-Stokes equations with perturbation in the following form:
\begin{align}
  \hat{M} \frac{\partial}{\partial t} \bd{q}(\bd{x}, t)
  = \bd{\cal F}[\bd{q}] + \epsilon \bd{E}(\bd{x}, t).
  \label{eq:q-E}
\end{align}
We assume that the perturbed solution is in the vicinity of the limit-cycle solution $\tilde{\bd{q}}(\bd{x}, \theta)$.
Using the adjoint zero eigenfunction $\tilde{\bd{Q}}^\ast(\bd{x}, \theta)$,
we project the dynamics of the perturbed equation~(\ref{eq:q-E})
onto the unperturbed limit-cycle solution to yield
\begin{align}
  \dot{\theta}(t)
  &= \Bigl\langle \tilde{\bd{Q}}^\ast(\bd{x}, \theta), \, 
  \hat{M} \frac{\partial}{\partial t} \bd{q}(\bd{x}, t) \Bigr\rangle
  \simeq \Bigl\langle \tilde{\bd{Q}}^\ast(\bd{x}, \theta), \, 
  \bd{\cal F}[\tilde{\bd{q}}] + \epsilon \bd{E}(\bd{x}, t) \Bigr\rangle
  \nonumber \\
  &= \omega_n + \epsilon \, \Bigl\langle \tilde{\bd{Q}}^\ast(\bd{x}, \theta), \, \bd{E}(\bd{x}, t) \Bigr\rangle
  = \omega_n + \epsilon \int_D \, \tilde{\bd{U}}^\ast(\bd{x}, \theta) \cdot \bd{K}(\bd{x}, t) \, d\bd{x},
  \label{eq:theta-E}
\end{align}
where we approximated $\bd{q}(\bd{x}, t)$ by the unperturbed limit-cycle solution $\tilde{\bd{q}}(\bd{x}, \theta)$,
and also used Eqs.~(\ref{eq:q-q0}), (\ref{eq:Q0-q0}), and (\ref{eq:normalization}).
The phase equation describing periodic flows under weak perturbation is then approximately obtained after the higher order terms are neglected as
\begin{align}
  \dot{\theta}(t)
  = \omega_n + \epsilon \int_D \, \bd{Z}(\bd{x}, \theta) \cdot \bd{K}(\bd{x}, t) \, d\bd{x}.
  \label{eq:phase}
\end{align}
As we define the phase sensitivity function $\bd{Z}(\bd{x}, \theta)$ with respect to the velocity fields,
we realize that
\begin{align}
  \bd{Z}(\bd{x}, \theta) = \tilde{\bd{U}}^\ast(\bd{x}, \theta),
\end{align}
from comparing Eqs.~(\ref{eq:theta-E}) and (\ref{eq:phase}).
Thus, the phase sensitivity function with respect to the velocity fields is the adjoint zero eigenfunction of velocity field and can be obtained by numerically solving Eqs.~(\ref{eq:U1ast-V1ast}) and (\ref{eq:P1ast}).
Using this adjoint-based framework, we can evaluate the spatial phase sensitivity fields using a single pair of forward and adjoint simulations.
This is unlike the traditional direct method, where the sensitivity field is obtained by solving the Navier-Stokes equations with added impulse perturbation at each grid point at each phase, thereby, requiring an $N_{grid}\times N_{phase}$ of simulations (where $N_{grid}$ is the number of grid points and $N_{phase}$ is the number of phases at which computation is required).

Let us examine the properties of the phase sensitivity function.
According to Eq.~(\ref{eq:P1ast}),
the phase sensitivity function is divergence-free, which can be expressed as
$\nabla \cdot \bd{Z}(\bd{x}, \theta) = 0$.
Here, we consider scalar-potential-based perturbations to the fluid velocity field as
$\bd{K}(\bd{x}, t) = - \nabla \Psi(\bd{x}, t)$,
where the scalar potential is denoted by $\Psi$ and the perturbation is designed such that $\Psi \rightarrow 0$ at the far-field.
In this case, the phase response becomes zero,
which implies that application of scalar-potential-based perturbations to the fluid velocity field does not affect the phase, i.e.,
\begin{align}
  \int_D \, \bd{Z}(\bd{x}, \theta) \cdot \bd{K}(\bd{x}, t) \, d\bd{x}
  = -\int_D \, \bd{Z}(\bd{x}, \theta) \cdot \Bigl[ \nabla \Psi(\bd{x}, t) \Bigr] \, d\bd{x}
  = \int_D \, \Bigl[ \nabla \cdot \bd{Z}(\bd{x}, \theta) \Bigr] \Psi(\bd{x}, t) \, d\bd{x}
  = 0.
  \label{eq:zero}
\end{align}

Next, let us consider the curl of the phase sensitivity function $\nabla \times \bd{Z}(\bd{x}, \theta)$.
Here, we consider vector-potential-based perturbations to the fluid velocity field as
$\bd{K}(\bd{x}, t) = \nabla \times \bd{A}(\bd{x}, t)$,
where the vector potential is denoted by $\bd{A}$ with $\bd{A} \rightarrow \bd{0}$ at the far-field.
In this case,
the phase response can be rewritten in the following form:
\begin{align}
  \int_D \, \bd{Z}(\bd{x}, \theta) \cdot \bd{K}(\bd{x}, t) \, d\bd{x}
  &= \int_D \, \bd{Z}(\bd{x}, \theta) \cdot \Bigl[ \nabla \times \bd{A}(\bd{x}, t) \Bigr] \, d\bd{x}
  = \int_D \, \Bigl[ \nabla \times \bd{Z}(\bd{x}, \theta) \Bigr] \cdot \bd{A}(\bd{x}, t) \, d\bd{x}.
  \label{eq:Omega}
\end{align}
This implies that the curl of the phase sensitivity function
quantifies the phase response to the vector potential $\bd{A}$.
However,
since the interpretation of phase sensitivity with respect to the perturbation in terms of the vector potential is challenging,
let us relate $\nabla \times \bd{Z}$ with the phase sensitivity for the vorticity, $\bd{Z}_{\bd{\omega}}$.
Here, consider a weak vortical perturbation $\epsilon \bd{K}_{\bd{\omega}}$ added to the vorticity transport equation as 
\begin{align}
  \frac{\partial}{\partial t} \bd{\omega}(\bd{x}, t)
  = -\bd{u} \cdot \nabla \bd{\omega}
  + \bd{\omega} \cdot \nabla \bd{u}
  + \rm Re^{-1} \nabla^2 \bd{\omega}
  + \epsilon \bd{K}_{\bd{\omega}}(\bd{x}, t).
\end{align}
Since we seek the periodic solution of the Navier-Stokes equation, both velocity and the vorticity are periodic in time.
This allows us to write
\begin{align}
  \dot{\theta}(t)
  &= \omega_n + \epsilon \int_D \, \bd{Z}(\bd{x}, \theta) \cdot \bd{K}(\bd{x}, t) \, d\bd{x}
  = \omega_n + \epsilon \int_D \, \bd{Z}_{\bd{\omega}}(\bd{x}, \theta) \cdot \bd{K}_{\bd{\omega}}(\bd{x}, t) \, d\bd{x}
  \nonumber \\
  &= \omega_n + \epsilon \int_D \, \Bigl[ \nabla \times \bd{Z}_{\bd{\omega}}(\bd{x}, \theta) \Bigr]
  \cdot \bd{K}(\bd{x}, t) \, d\bd{x}.
  \label{eq:phase_vort}
\end{align}
Hence, $\bd{Z}$ and $\bd{Z}_{\bd{\omega}}$ are related as
$\bd{Z} = \nabla \times \bd{Z}_{\bd{\omega}}$.
It follows that
$\nabla \times \nabla \times \bd{Z}_{\bd{\omega}} = \nabla \times \bd{Z}$.
Hence, we obtain a Poisson relation between the curl of phase sensitivity and phase sensitivity in terms of vorticity as
\begin{align}
  \nabla^2 \bd{Z}_{\bd{\omega}} = -\nabla \times \bd{Z},
\end{align}
and $\nabla \cdot \bd{Z}_{\bd{\omega}}=0$ is applied without the loss of generality.
Thus,
a positive (negative) value of $\nabla \times \bd{Z}$ would imply
a positive (negative) source for the Poisson equation of $\bd{Z}_{\bd{\omega}}$,
hence results in a local increment (drop) of $\bd{Z}_{\bd{\omega}}$.

Establishing the phase sensitivity function enables us to consider the forced synchronization of periodic flows.
Consider the perturbation to be separable, i.e., $\bd{K}(\bd{x}, t) = \bd{a}(\bd{x}) b(t)$,
the phase equation can be written as $\dot{\theta}(t) = \omega_n + \epsilon \zeta(\theta) b(t)$.
The effective phase sensitivity function is given by
\begin{align}
  \zeta(\theta)
  = \int_D \, \bd{Z}(\bd{x}, \theta) \cdot \bd{a}(\bd{x}) \, d\bd{x}.
  \label{eq:zeta}
\end{align}
When the perturbation takes a form of periodic forcing, i.e., $b(t) = b_{\rm p}(\omega_f t)$ with $\omega_f \sim \omega_n$,
the phase equation can be written as $\dot{\theta}(t) = \omega_n + \epsilon \zeta(\theta) b_{\rm p}(\omega_f t)$.
In this case, we can introduce a slow phase variable as $\theta(t) = \omega_f t + \psi(t)$. We can then rewrite the phase equation as
$\dot{\psi}(t) = \omega_n - \omega_f + \epsilon \zeta(\omega_f t + \psi) b_{\rm p}(\omega_f t)$. Because the frequency mismatch, $\omega_n - \omega_f$, and the intensity of the periodic forcing, $\epsilon$, are small, the dynamics of the phase variable $\psi$ becomes slow.
We can thus apply the averaging over the period to obtain the following phase equation:
\begin{align}
  \dot{\psi}(t)
  = \omega_n - \omega_f + \epsilon \Gamma_{\rm p}(\psi),
  \label{eq:psi-Gamma}
\end{align}
where the phase coupling function is defined as
\begin{align}
  \Gamma_{\rm p}(\psi)
  \equiv \frac{1}{2\pi} \int_0^{2\pi} \, \zeta(\lambda + \psi) b_{\rm p}(\lambda) \, d\lambda.
  \label{eq:Gamma}
\end{align}
The phase coupling and effective phase sensitivity functions could be used
for analyzing the forced synchronization characteristics, lock-in behavior of periodic flows to external forcing. As an application,
the forced synchronization of vortex shedding for a circular cylinder to periodic external forcing have been studied in Refs.~\cite{ref:taira18,ref:khodkar20}.

\subsection{Adjoint formulation for the immersed boundary projection method}

We provided the phase-based description of periodic flows using the adjoint formulation of the Navier-Stokes equations.
Any well-established numerical scheme can be used to perform the phase based analysis for periodic flows through the present formulation.
One such scheme, the immersed boundary projection method can be used to analyze the flows over bluff bodies of arbitrary shapes.

Here, let us demonstrate the implementation of the adjoint method to the immersed boundary method to solve the Navier-Stokes equations.
In this method, a body $B$ can is generated using the introduction of boundary forces along the surface.
The no-slip boundary condition are enforced using the discrete delta functions.
The incompressible Navier-Stokes equations in an immersed boundary projection method can be expressed as
\begin{align}
  \frac{\partial}{\partial t} \bd{u}(\bd{x}, t)
  &= - \bd{u} \cdot \nabla \bd{u} - \nabla p + {\rm Re}^{-1} \nabla^2 \bd{u}
  + \int_{\partial B} \, \bd{f}(\bd{\xi}, t) \delta(\bd{\xi} - \bd{x}) \, d\bd{\xi},
  \label{eq:u-v_ibpm} \\
  0
  &= \nabla \cdot \bd{u},
  \label{eq:p_ibpm} \\
  \bd{0}
  &= \int_D \, \bd{u}(\bd{x}, t) \delta(\bd{x} - \bd{\xi}) \, d\bd{x}.
  \label{eq:f-g_ibpm}
\end{align}
The spatial variables are defined as $\bd{x} \in D$ and $\bd{\xi} \in \partial B$.
Here, we assume the immersed surface $\partial B$ is stationary without the loss of generality.
The variable $\bd{q}(\bd{x}, t)$ can be modified to include the boundary forces $\bd{f}$ as $\bd{q}(\bd{x}, t) = (\bd{u}, \, p, \, \bd{f})^{\rm T}$.
Hence, the set of Eqs.~(\ref{eq:u-v_ibpm}), (\ref{eq:p_ibpm}), and (\ref{eq:f-g_ibpm}) can be written as
\begin{align}
  \hat{M} \frac{\partial}{\partial t} \bd{q}(\bd{x}, t)
  = \bd{\cal F}[\bd{q}],
  \label{eq:q_ibpm}
\end{align}
where $\hat{M} = {\rm diag}(1, \, 1, \, 1, \, 0, \, 0, \, 0, \,0)$ and the right-hand-side term $\bd{\cal F}[\bd{q}]$ now includes the boundary forces and the no-slip boundary condition along the immersed surface.
The components in the linear operator in Eqs.~(\ref{eq:calL}) and (\ref{eq:calJ}) can be modified to include the additional terms as
\begin{align}
  \hat{\cal J} \bd{q}^\prime = 
  \begin{cases}
    \displaystyle
    - \bd{u}^\prime \cdot \nabla \tilde{\bd{u}}
    - \tilde{\bd{u}} \cdot \nabla \bd{u}^\prime
    - \nabla p^\prime
    + {\rm Re}^{-1} \nabla^2 \bd{u}^\prime
    + \int_{\partial B} \, \bd{f}^\prime(\bd{\xi}, \theta) \delta(\bd{\xi} - \bd{x}) \, d\bd{\xi},
    \\[3mm]
    \displaystyle
    \nabla \cdot \bd{u}^\prime,
    \\[3mm]
    \displaystyle
    \int_D \, \bd{u}^\prime(\bd{x}, \theta) \delta(\bd{x} - \bd{\xi})\,d\bd{x}.
  \end{cases}
  \label{eq:calJ_ibpm}
\end{align}
The inner product defined in Eq.~(\ref{eq:inner-product}) can also be modified to include the boundary forces as
\begin{align}
  \Bigl\langle {\bd{q}^\prime}^\ast(\bd{x}, \theta), \, \bd{q}^\prime(\bd{x}, \theta) \Bigr\rangle = 
  \int_D \,
  \Bigl[ {\bd{u}^\prime}^\ast(\bd{x}, \theta) \cdot \bd{u}^\prime(\bd{x}, \theta) + {p^\prime}^\ast(\bd{x}, \theta) p^\prime(\bd{x}, \theta) \Bigr] \,
  d\bd{x}
  + \int_{\partial B} \,
  \Bigl[ {\bd{f}^\prime}^\ast(\bd{\xi}, \theta) \cdot \bd{f}^\prime(\bd{\xi}, \theta) \Bigr] \, d\bd{\xi}.
  \label{eq:inner-product_ibpm}
\end{align}
The components in the linearized adjoint operator in Eqs.~(\ref{eq:calLast}) and (\ref{eq:calJast}) are also modified as
\begin{align}
  \hat{\cal J}^\ast \bd{q}_1^\ast = 
  \begin{cases}
    \displaystyle
    - {u_{x}^\prime}^\ast \nabla \tilde{u}_{x}
    - {u_{y}^\prime}^\ast \nabla \tilde{u}_{y}
    - {u_{z}^\prime}^\ast \nabla \tilde{u}_{z}
    + \tilde{\bd{u}} \cdot \nabla {\bd{u}^\prime}^\ast
    - \nabla {p^\prime}^\ast
    + {\rm Re}^{-1} \nabla^2 {\bd{u}^\prime}^\ast
    + \int_{\partial B}  \, {\bd{f}^\prime}^\ast(\bd{\xi}, \theta) \delta(\bd{\xi} - \bd{x})\, d\bd{\xi},
    \\[3mm]
    \displaystyle
    \nabla \cdot {\bd{u}^\prime}^\ast,
    \\[3mm]
    \displaystyle
    \int_D  \, {\bd{u}^\prime}^\ast(\bd{x}, \theta) \delta(\bd{x} - \bd{\xi})\,d\bd{x}.
  \end{cases}
  \label{eq:calJast_ibpm}
\end{align}
Moreover, the adjoint equations, Eqs.~(\ref{eq:U1ast-V1ast}) and (\ref{eq:P1ast}), become
\begin{align}
  \frac{\partial}{\partial s} \tilde{\bd{U}}^\ast(\bd{x}, -\omega_n s)
  &=
  - {\tilde{U}_{x}}^\ast \nabla \tilde{u}_{x}
  - \tilde{U}_{y}^\ast \nabla \tilde{u}_{y}
  - \tilde{U}_{z}^\ast \nabla \tilde{u}_{z}
  + \tilde{\bd{u}} \cdot \nabla \tilde{\bd{U}}^\ast
  - \nabla \tilde{P}^\ast
  + {\rm Re}^{-1} \nabla^2 \tilde{\bd{U}}^\ast
  + \int_{\partial B} \, \tilde{\bd{F}}^\ast(\bd{\xi}, -\omega_n s) \delta(\bd{\xi} - \bd{x}) \, d\bd{\xi},
  \label{eq:U1ast-V1ast_ibpm} \\
  0
  &= \nabla \cdot \tilde{\bd{U}}^\ast,
  \label{eq:P1ast_ibpm} \\
  \bd{0}
  &= \int_D \, \tilde{\bd{U}}^\ast(\bd{x}, -\omega_n s) \delta(\bd{x} - \bd{\xi}) \, d\bd{x}.
  \label{eq:F1ast-G1ast_ibpm}
\end{align}
The inclusion of the boundary forces and the enforcement of no-slip boundary condition on the surface
does not modify the original adjoint formulation and the definitions of the phase sensitivity functions remain the same. 
In this study, we demonstrate the capability of adjoint formulation of the immersed boundary projection scheme
by performing the phase sensitivity analysis over circular cylinder and symmetric airfoils of different thicknesses at high angles of attack.

\section{Results} \label{sec:3}

In this section, the adjoint-based phase-reduction analysis is applied to time-periodic flows.
The present approach is first validated with a two-dimensional time-periodic laminar flow over a circular cylinder.
The present results from this cylinder wake analysis are compared to those from a prior impulse response-based phase-reduction analysis~\cite{ref:khodkar20}.
Next, we utilize the adjoint-based phase-reduction analysis to reveal the phase dynamic properties for time-periodic separated flows over canonical airfoils over a range of angles of attack.
With the immersed boundary formulation being incorporated into the present approach, the phase response properties of flows over bodies of arbitrary surface geometries can be uncovered efficiently.

\subsection{Boundary conditions for the flow fields and the adjoint equations} \label{sec:boundary}

For the present work,
two-dimensional periodic flows over a circular cylinder and symmetric airfoils simulated by the immersed boundary formulation~\cite{ref:taira07,ref:taira08,ref:taira17} are considered.
For these simulations, the inflow and far-field boundary conditions are prescribed as
$\bd{u} = \bd{u}_{\infty} = (u_{\infty}, \, v_{\infty})^{\rm T} = (1, \, 0)^{\rm T}$.
Along the outlet, we prescribe the convective outflow condition as
$( \partial_t + u_{\infty} \, \partial_x ) \bd{u}(\bd{x}, t) = \bd{0}$.

The components of $\tilde{\bd{q}}(\bd{x}, \theta)$ satisfy the same conditions as $\bd{u}$,
whereas the components of ${\bd{q}^\prime}(\bd{x}, \theta)$ satisfy the Dirichlet-zero boundary conditions of
${\bd{u}^\prime}(\bd{x}, \theta)|_{\partial D} = \bd{0}$
except for the outlet, which prescribes
$\partial_x {\bd{u}^\prime}(\bd{x}, \theta) = \bd{0}, \, {p^\prime}(\bd{x}, \theta) = 0$.
The components of ${\bd{q}^\prime}^\ast(\bd{x}, \theta)$ satisfy the Dirichlet-zero boundary conditions of
${\bd{u}^\prime}^\ast(\bd{x}, \theta)|_{\partial D} = \bd{0}$
except for the outlet, which prescribes
$( {\rm Re}^{-1} \, \partial_x + \tilde{u}(\bd{x}, \theta) ) {\bd{u}^\prime}^\ast(\bd{x}, \theta) = \bd{0}, \, {p^\prime}^\ast(\bd{x}, \theta) = 0$.

\begin{table}
    \centering
    \begin{tabular}{lccc} \hline
                                                & $C_L$         & $C_D$             & $St$  \\ \hline \hline
    Present                                     & $\pm 0.338$   & $1.366\pm 0.009$  & 0.165 \\
    Taira \& Nakao (2018) \cite{ref:taira18}    & $\pm 0.328$   & $1.35 \pm 0.009$  & 0.165 \\
    Liu, Zheng \& Sung (1998) \cite{ref:liu98}  & $\pm 0.339$   & $1.35 \pm 0.012$  & 0.165 \\
    Linnick \& Fasel (2005) \cite{ref:linnick05}& $\pm 0.337$   & $1.34 \pm 0.009$  & 0.165 \\
    Canuto \& Taira (2015) \cite{ref:taira15}   & $\pm 0.329$   & $1.34 \pm 0.009$  & 0.167 \\ \hline
    \end{tabular}
    \caption{Comparison of forces ($C_L$, $C_D$) and Strouhal number ($St$) for flow over a circular cylinder at $Re = 100$.}
    \label{table:cylinder}
\end{table}

\begin{figure}
   \centering
   \includegraphics[width=0.9\textwidth]{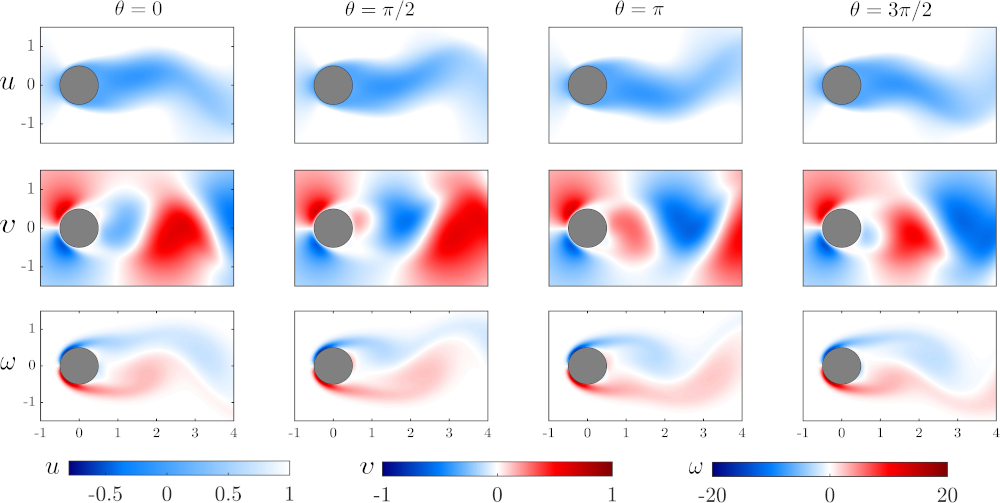}
    \caption{Streamwise velocity $u$, transverse velocity $v$ and vorticity $\omega$ fields at phases of $\theta = [0,~\pi/2,~\pi,~3\pi/2]$ for flow over a circular cylinder at $Re = 100$.}
    \label{fig:cylinder_flowfield}
\end{figure}

\begin{figure}
    \centering
    \includegraphics[width=0.9\textwidth]{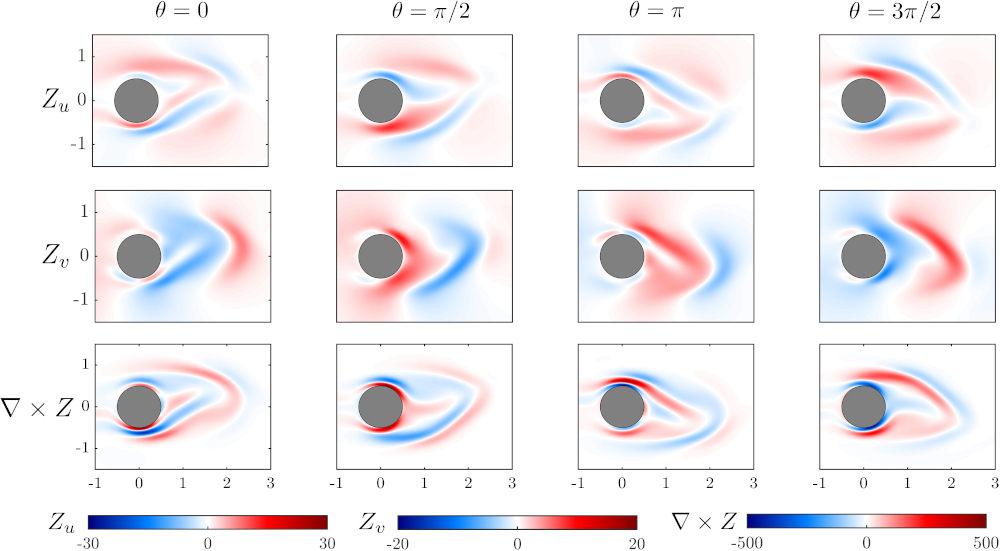}
    \caption{Phase-sensitivity function in terms of velocity ($Z_u$, $Z_v$) and vector potential ($\nabla \times \bd{Z}$) at phases $\theta = [0,~\pi/2,~\pi,~3\pi/2]$ for flow over a circular cylinder at $Re = 100$.}
    \label{fig:cyl_phsensitivity}
\end{figure}
\subsection{Circular Cylinder Wake}

First, we examine the phase dynamic characteristics of the two-dimensional incompressible laminar periodic flows over a circular cylinder at a diameter-based Reynolds number of $Re = 100$.  This flow is selected to validate the present approach against an impulse response-based approach~\cite{ref:khodkar20}, which requires a significant amount of computational effort. To initialize the analysis, we seek the time-periodic wake (limit cycle) through direct numerical simulation.  This is achieved by performing the forward simulation that solves the incompressible Navier-Stokes equations, Eqs.~(\ref{eq:u-v_ibpm}), (\ref{eq:p_ibpm}), and (\ref{eq:f-g_ibpm}).  In this example, we obtain the time-periodic von Karman vortex street that forms behind the circular cylinder.  At this Reynolds number, the laminar wake is unsteady with vortices shedding in an alternating manner from the top and bottom of the cylinder.  The simulation here is performed with the immersed boundary projection method~\cite{ref:taira07,ref:taira17}.  The velocity and pressure fields are discretized on a staggered Cartesian grid and the circular cylinder is generated in the flow field through the introduction of boundary forces along the cylinder surface, which is represented by a set of Lagrangian points.  The computational technique is formally second-order in space and time with first-order spatial accuracy near the cylinder surface, where discrete delta functions are used to enable the enforcement of the no-slip boundary condition.  

The computational domain is chosen to be $\mathcal{D} = \{ (x, y) \in [-16, 16] \times [-30, 30] \}$ with the circular cylinder centered at the origin.  Because the current study requires both forward and adjoint simulations to be performed, the spatial domain is discretized with fine grids both upstream and downstream of the circular cylinder.  All spatial variables are non-dimensionalized with the cylinder diameter $d$ and the time scales are normalized by the convective time of $d/u_\infty$, where $u_{\infty}$ is the free stream velocity.  The smallest grid size is set to $\Delta x = 0.03$, and the time step is chosen to be $\Delta t = 0.01$ such that it satisfies the CFL condition of $u_\infty \Delta t/\Delta x_\text{min} < 0.33$.  For the present simulation, Dirichlet conditions are specified along the inlet and far-field boundaries to match free stream velocity profile of $(U_\infty, 0)$.  For the forward simulations, the convective outflow boundary condition of $\partial\boldsymbol{u}/\partial t + u_\infty \partial\boldsymbol{u}/\partial x = 0$ is used for the outlet.  On the other hand, for the adjoint simulation, Neumann boundary condition is prescribed along the adjoint outlet.  

The time-periodic von Karman vortex shedding wake is obtained from the forward simulation.  The velocity and vorticity fields are shown in figure~\ref{fig:cylinder_flowfield}, which agrees well with those reported in past studies~\cite{ref:taira07,ref:taira13,ref:taira15}.  Here, the flow fields are shown for phases of $\theta = [0,~\pi/2,~\pi,~3\pi/2]$; the phase of $\theta = 0$ corresponds to $C_L = 0$ and $\dot{C}_L = \max \dot{C}_L$.  Note that the flow fields offset by a phase of $\pi$ are symmetric about the $y=0$ axis.  The vortices that shed periodically imposes unsteady forces onto the cylinder, which results in sinusoidal lift and drag forces over time.  The lift and drag coefficients $(C_L, C_D) \equiv (F_L, F_D)/(\frac{1}{2}\rho_\infty u_\infty^2 d)$ along with the Strouhal number $St \equiv f d/u_\infty$ are in agreement with those from previous studies, as summarized in Table~\ref{table:cylinder}.  

With the baseline time-periodic wake obtained, the adjoint simulation is performed to determine the phase-sensitivity function over the spatial domain.  The spatial profiles of the phase-sensitivity function in terms of $Z_u$, $Z_v$, and $\nabla \times \bd{Z}$ at $\theta = [0,~\pi/2,~\pi,~3\pi/2]$ are presented in figure~\ref{fig:cyl_phsensitivity}.  As the adjoint simulation perform time integration backwards in time, we observe that the phase-sensitivity functions $Z_u$, $Z_v$, and $\nabla \times \bd{Z}$ advect in the opposite direction from the flow field shown in figure~\ref{fig:cylinder_flowfield}.  In contrast to the forward simulation, the phase sensitivity functions $Z_u$, $Z_v$, and $\nabla \times \bd{Z}$ exhibit large-amplitude fluctuations in the aft side of the cylinder and in the boundary layer region.  This suggests that the phase can be influenced efficiently in the wake region and along the boundary layer near the separation point.  It should be reminded that the profiles of $Z_u$ and $Z_v$ are related to $\nabla \times \bd{Z}$ through the curl operator.  

Due to the $2\pi$ periodicity of the flow field, for a two-dimensional flow, the flow field variable satisfy
\begin{align}
  \tilde{u}(x, y, \theta + \pi) &=  \tilde{u}(x, -y, \theta), \\
  \tilde{v}(x, y, \theta + \pi) &= -\tilde{v}(x, -y, \theta), \\
   \omega_z(x, y, \theta + \pi) &= - \omega_z(x, -y, \theta). 
  \label{eq:uvw_cylinder_symmetry}
\end{align}
The phase sensitivity fields follow that 
\begin{align}
  Z_u(x, y, \theta + \pi) &=  Z_u(x, -y, \theta), \\
  Z_v(x, y, \theta + \pi) &= -Z_v(x, -y, \theta), \\
  \nabla \times \bd{Z}(x, y, \theta + \pi) &= -\nabla \times \bd{Z}(x, -y, \theta).
  \label{eq:Zuvw_cylinder_symmetry}
\end{align}
Since the base flow is symmetric about the $x$-axis with a phase shift of $\pi$, the phase sensitivity functions $Z_u$, $Z_v$, and $\nabla \times \bd{Z}$ exhibit the same properties.  The phase sensitivity functions are in excellent agreement with those reported by Khodkar and Taira~\cite{ref:khodkar20} with enhanced spatial resolution and fidelity.  Because the present approach simulated $Z_u$ and $Z_v$ through the adjoint code, the necessary computational resource is significantly reduced compared to the impulse based approach~\cite{ref:khodkar20}.  We note that the phase sensitivity function in Khodkar and Taira~\cite{ref:khodkar20} is defined with a sign difference.

\begin{figure}
    \centering
    \includegraphics[width=0.7\textwidth]{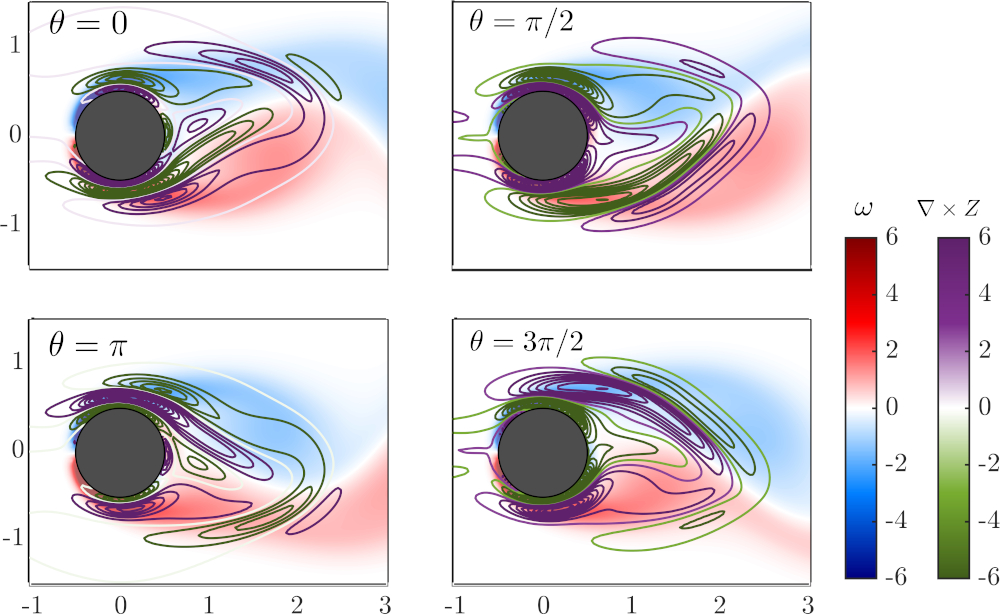}
    \caption{Phase-sensitivity function $\nabla \times \bd{Z}$ (line contour) superimposed on the vorticity fields ($\omega$; filled contour) at phases $\theta = [0,~\pi/2,~\pi,~3\pi/2]$ for flow over a circular cylinder at $Re = 100$.}
    \label{fig:cyl_Zw_vort}
\end{figure}

What is strikingly different about the phase sensitivity functions from the base flow is that the profiles show finer layer-like structures appearing in the aft region of cylinder (right side).  We note that since $\nabla \times \bd{Z}$ acts as a source for $\bd{Z}_{\bd{\omega}}$ through the Poisson equation, a positive value of $\nabla \times \bd{Z}$ would correspond to local increment in $\bd{Z}_{\bd{\omega}}$ and a negative value of $\nabla \times \bd{Z}$ would correspond to local drop in $\bd{Z}_{\bd{\omega}}$.  It also follows that the phase sensitivity fields $\nabla \times \bd{Z}$ and $\bd{Z}_{\bd{\omega}}$ encodes similar behavior, although $\nabla \times \bd{Z}$ field would result in compact structures.  Now, let us superpose the phase sensitivity function $\nabla \times \bd{Z}$ on the vorticity field $\omega$ in figure~\ref{fig:cyl_Zw_vort} to take a detailed look.  As we examine the vorticity and $\nabla \times \bd{Z}$ fields at $\theta = 0,\,\pi/2$, we notice that positive layers of $\nabla \times \bd{Z}$ are adjacent to top and bottom surfaces of the cylinder, near the separation points.  This indicates through the Poisson equation that the injection of positive vorticity leads to phase advancement in these regions.  
At these particular phases, the top positive surface layer of $\nabla \times \bd{Z}$ highlights the regions sensitive to phase advancement, translating to accelerated vortex shedding.  The bottom positive surface layer of $\nabla \times \bd{Z}$ identifies the region also promoting phase advancement but through the promotion of the formation of the bottom vortex.  Through Eqs.~(\ref{eq:uvw_cylinder_symmetry}), (\ref{eq:Zuvw_cylinder_symmetry}), we observe that an opposite actuation would result in phase advancement for $\theta=\pi,\,3\pi/2$.  
Also noteworthy here is that there is a negative layer of $\nabla \times \bd{Z}$ directly outside of the positive layer.  This suggests that the radial location to which vorticity is added is important in changing the phase of the shedding process.  

\begin{figure}
\centering
  \includegraphics[width=0.85\textwidth]{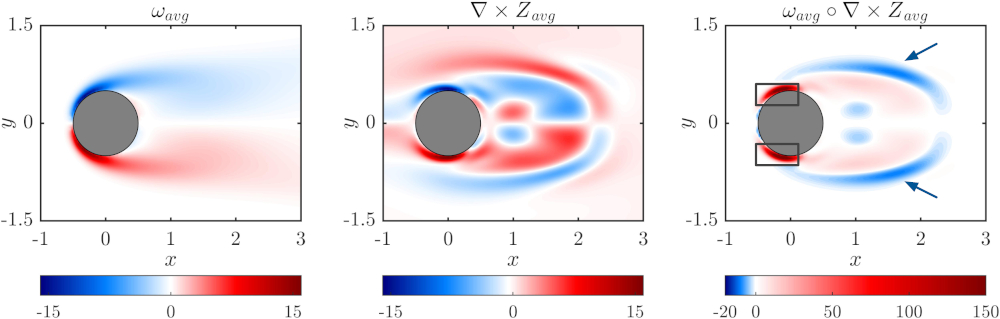}
  \caption{The phase averaged vorticity ($\omega$), the phase-sensitivity function in terms of vector potential ($\nabla \times Z$), and the Hadamard product ($\omega \circ \nabla \times \bd{Z}$) for flow over a circular cylinder at $Re = 100$.}
  \label{fig:cyl_Zw_dotw_phaseavg}
\end{figure}

Let us also consider the phase-averaged vorticity and phase sensitivity fields in figure~\ref{fig:cyl_Zw_dotw_phaseavg}.  We can observe that the averaged phase sensitivity field $\nabla \times \bd{Z}$ has highly sensitive regions where the boundary layer develops over the cylinder and in the wake region near $x/d \approx 2$.  In order to assess whether these regions align with the vorticity field, we take the correlation (Hadamard product) of $\omega$ and $\nabla \times \bd{Z}$ are shown in figure~\ref{fig:cyl_Zw_dotw_phaseavg} (right).  We note that this product $\omega \circ \nabla \times \bd{Z}$ looks qualitatively similar to the global-sensitivity fields, as reported in Ref.~\cite{ref:luchini14,ref:giannetti07}.  The phase-averaged phase sensitivity field highlights the regions that are largely sensitive for perturbations.  A high correlation between $\nabla \times \bd{Z}$ and $\omega$ is observed near the flow separation locations and the regions with high shear in the flow field, as highlighted in figure~\ref{fig:cyl_Zw_dotw_phaseavg} (right).  This indicates that phase-sensitivity analysis has the potential to identify sensitive regions that are relevant to the flow physics.

\subsection{High-Incidence Airfoil Wake}

In this section, let us investigate the phase dynamic characteristics of incompressible laminar periodic flows over symmetric NACA airfoils of different thicknesses at post stall angles of attack at a chord-based Reynolds number of $Re = 100$.  In particular, we examine the influence of angle of attack and thickness on the phase sensitivity fields obtained from the adjoint-based phase reduction approach.  

\begin{figure}
    \centering
    \includegraphics[width=0.8\textwidth]{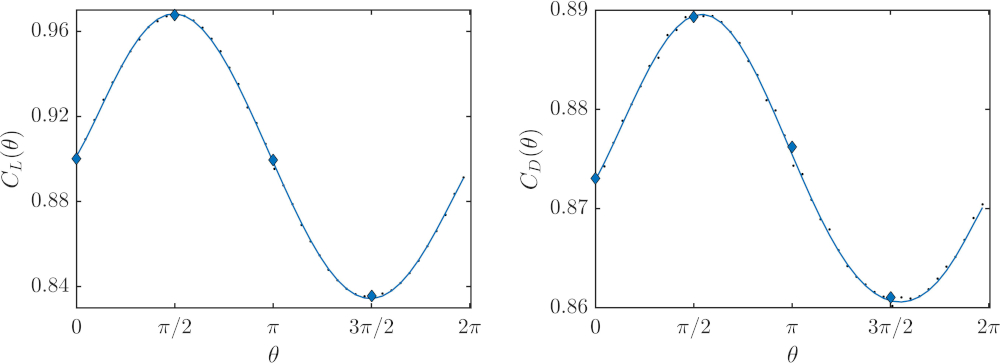}
    \caption{The lift and drag coefficients as function of phase for NACA0012 airfoil at $\alpha = 35^\circ$ for $Re = 100$.}
    \label{fig:airfoil_lift}
\end{figure}

\begin{figure}
    \centering
    \includegraphics[width=0.85\textwidth]{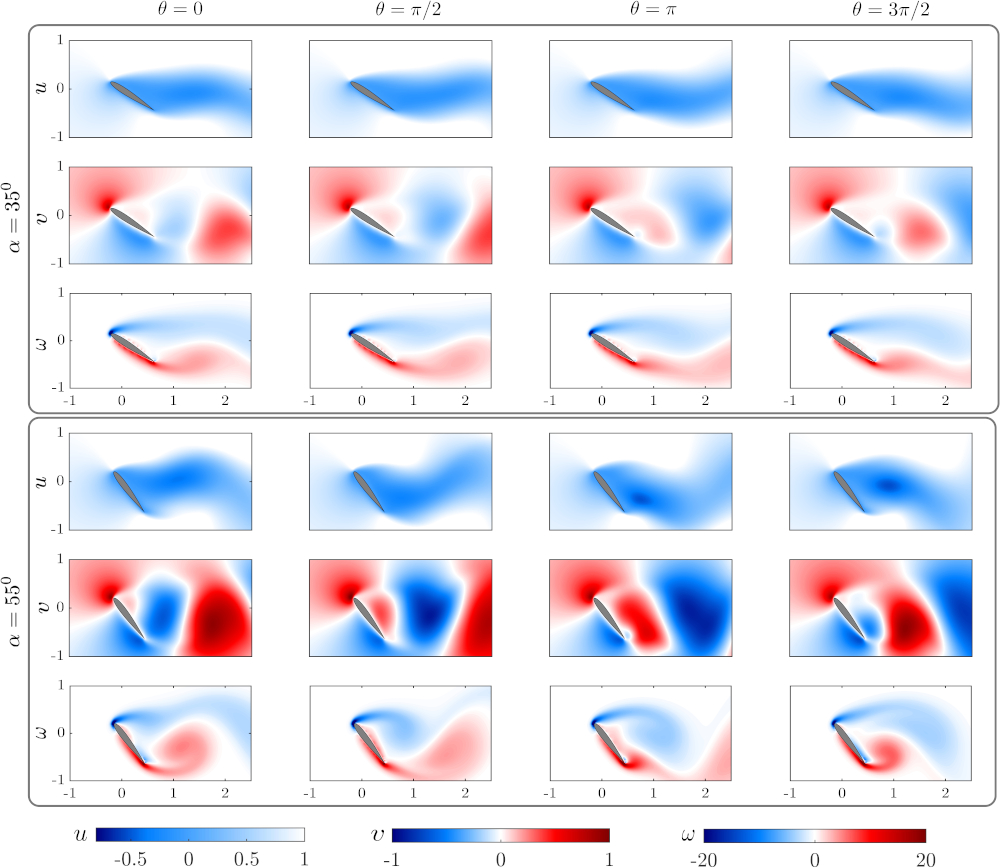}
    \caption{The velocity and vorticity fields at phases $\theta = [0,~\pi/2,~\pi,~3\pi/2]$ for flows over NACA0012 airfoil at angles of attack $35^\circ$ and $55^\circ$ for $Re = 100$.}
    \label{fig:uvw_airfoil}
\end{figure}

The time-periodic wakes of airfoils at post-stall angles of attack are obtained through the direct numerical simulations.  The computational setup is similar to the cylinder wake simulation.  The laminar flow over various airfoils are computed using the immersed boundary projection method~\cite{ref:taira07, ref:taira08, ref:taira17} where these airfoils of different thicknesses and angles of attack are generated in the Cartesian flow field through the introduction of boundary forces along their surface.  The computational domain is chosen to be $\mathcal{D} = \{(x, y) \, \in \, [-16, 16] \times [-30, 30]\}$ with the quarter-chord of the airfoil placed at the origin.  All spatial variables are non-dimensionalized with the airfoil chord length $c$ and the time scales are normalized by the convective time of $c/u_\infty$, where $u_{\infty}$ is the free stream velocity. The smallest grid size is set to $\Delta x = 0.02$, and the time step is chosen to be $\Delta t = 0.005$ such that it satisfies the CFL condition of $u_\infty \Delta t/\Delta x_{\min} < 0.4$.  For the forward simulation, Dirichlet conditions are specified along the inlet and far-field boundaries to match free stream velocity profile of $(u_\infty, 0)$, and the convective outflow boundary condition of $\partial \mathbf{u}/\partial t + u_\infty\partial\mathbf{u}/\partial x = 0$ is used for the outlet. For the adjoint simulation, Dirichlet boundary conditions are prescribed for all the boundaries except the outlet and the conditions $({\rm Re}^{-1}\partial_x+\tilde{u}(\bd{x},\theta)){\bd{u}^\prime}^\ast(\bd{x},\theta)=0,\,{p^{\prime}}^\ast(\bd{x},\theta)=0$ are prescribed at the adjoint outlet.
\begin{figure}[t]
    \centering
    \includegraphics[width=0.9\textwidth]{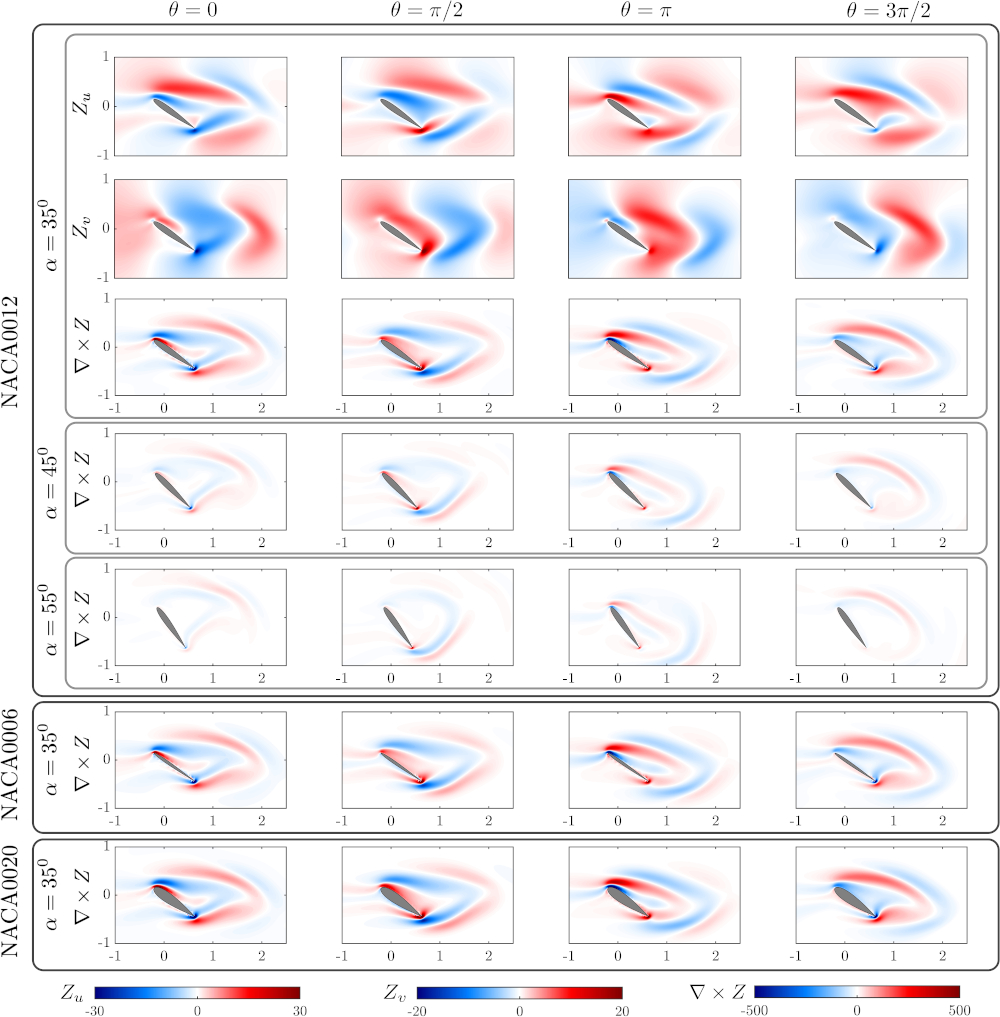}
    \caption{Phase sensitivity functions in terms of velocity ($Z_u$, $Z_v$) and vector potential ($\nabla \times \bd{Z}$) at various phases for flows over NACA0012 airfoil at $\alpha=35^\circ$, $45^\circ$, $55^\circ$. Also shown are $\nabla \times \bd{Z}$ fields for NACA0006 and NACA0020 airfoils at $\alpha = 35^\circ$ for $Re = 100$.}
    \label{fig:Zuvw_airfoil}
\end{figure}
The flow around symmetric NACA airfoils with thicknesses ranging from $6\%$ to $24\%$ at various angles of attack from $35^\circ$ to $60^\circ$ are considered.  At these angles of attack, we observe unsteady time-periodic vortex shedding over the airfoils.  This is reflected as the limit-cycle oscillations of the lift and drag coefficients.  The phase ($\theta$) is defined based on the $C_L - \dot{C}_L$ plane, where $\theta = 0$ corresponds to $C_L = {\rm avg}\,C_{L}$ and $\theta = \pi/2$ corresponds to $C_L = \max{C_L}$.  The variation of lift and drag coefficients with respect to the phase for a NACA0012 airfoil at $\alpha = 35^\circ$ is shown in figure~\ref{fig:airfoil_lift}.  The velocity and vorticity fields over a NACA0012 airfoil at $\alpha = 35^\circ$ and $\alpha = 55^\circ$ for phases of $\theta = 0$, $\pi/2$, $\pi$, $3\pi/2$ are shown in figure~\ref{fig:uvw_airfoil}.  We observe large coherent structures in the separated regions over the airfoil from the higher velocity ($u,\,v$) and vorticity ($\omega$) fields at $\alpha = 55^\circ$ than those observed at $\alpha = 35^\circ$.  

Once the time-periodic wake is obtained from the forward simulation, the adjoint simulation is performed to obtain the high-fidelity phase sensitivity fields. The spatial profiles of phase sensitivity functions in terms of $Z_u$, $Z_v$, $\nabla \times \bd{Z}$ for NACA0012 airfoil at $\alpha = 35^\circ$ for representative phases $\theta = 0$, $\pi/2$, $\pi$, $3\pi/2$ are presented in the top three rows of figure~\ref{fig:Zuvw_airfoil}.  The spatial patterns in the phase sensitivity fields are qualitatively similar for airfoils of different thicknesses and angles of attack.  Similar to the case of cylinder, we observe streak-like patterns around the airfoil in the $\nabla \times \bd{Z}$ fields with alternating positive and negative distributions.  However, unlike the phase sensitivity fields of cylinder, large magnitudes of phase sensitivity appear at the leading and trailing edges of the airfoil.  This behavior of large concentrated magnitudes at the leading and trailing edges of the airfoils is seen in all the phase sensitivity fields $Z_u$, $Z_v$, $\nabla \times \bd{Z}$.  A high magnitude of phase sensitivity indicates that a small perturbation is sufficient to cause significant phase modification near edges of the wing.  This feature is consistent with leading and trailing edge regions being very influential to the wake as sources of vorticity.  Since, the leading and trailing edges of the airfoil are the distinct features with high curvature, the phase sensitivity function assumes maxima near these regions.  Similar to the case of cylinder, regions near separation are highlighted in $\nabla \times \bd{Z}$ fields for airfoils.  Consider $\nabla \times \bd{Z}$ for NACA0012 at $\alpha = 35^\circ$ shown in figure~\ref{fig:Zuvw_airfoil} (third row), the streak-like patterns with large magnitudes of $\nabla \times \bd{Z}$ follows the pressure side of the airfoil until leading edge separation occurs.

Furthermore, we examine the influence of the angle of attack and thickness on the phase sensitivity fields.  To study the effect of angle of attack, we compare the phase-sensitivity function in terms of the vector potential ($\nabla \times \bd{Z}$) of flows over NACA0012 at $\alpha = 35^\circ$, $45^\circ$ and $55^\circ$ as shown in the third, fourth and fifth rows of the figure~\ref{fig:Zuvw_airfoil}.  As the angle of attack increases, the magnitude of $\nabla \times \bd{Z}$ decrease drastically, especially when the angle of attack is changed from $\alpha = 35^\circ$ to $45^\circ$.  At the higher angles of attack, regions with high magnitudes of phase sensitivity are concentrated more compactly at the leading and trailing edge of the airfoils and even in the high shear regions of airfoil wake unlike $\alpha = 35^\circ$.  A similar trend is observed the phase sensitivity fields in terms of the velocity (not shown for brevity).  This behavior is a result of earlier separation and higher unsteadiness with large vortical structures in the airfoil wake at high angles of attack.  This means that a stronger perturbation and control effort is required to cause a modification to the flow as the angle of attack is increased.  

Next, we analyze the influence of thickness on the phase sensitivity fields.  To this end, we compare the $\nabla \times \bd{Z}$ fields of NACA0006, NACA0012, and NACA0020 at $\alpha = 35^\circ$ as shown in the third and the bottom two rows of the figure~\ref{fig:Zuvw_airfoil}.  The influence of thickness on the phase sensitivity fields is not as drastic as the effect of angle of attack. However, we observe some important changes in the spatial phase sensitivity profiles due to the thickness.  As the thickness is increased, the magnitude of phase sensitivity increases around the airfoil.  Additionally, we observe larger concentrated regions of high magnitudes of phase sensitivity especially around the leading edge of a thicker airfoil.  This is a consequence of lower curvature around the leading edge of thick airfoils, making it feasible to perturb the flow more easily when compared to a thin airfoil with high curvature.  

\begin{figure}[t]
    \centering
    \includegraphics[width=0.7\textwidth]{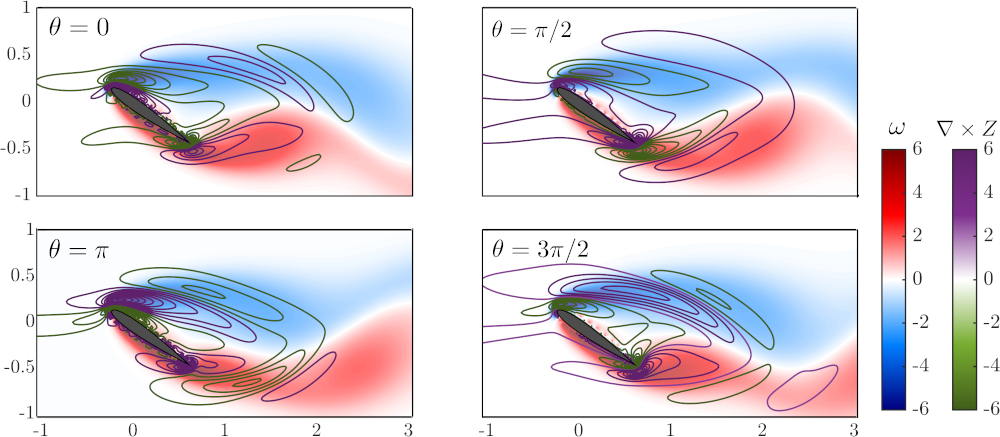}
    \caption{Iso-contours of phase sensitivity function in terms of vector potential ($\nabla \times \bd{Z}$) superposed on vorticity fields ($\omega$) at various phases for flow over NACA0012 airfoil at $\alpha = 35^\circ$ for $Re = 100$.}
    \label{fig:Zwvort_airfoil_lift}
\end{figure}

In addition to examining the influence of angle of attack and thickness, as the phase is defined based on the lift coefficient, phase modification through vortex shedding dynamics can be utilized for enhancement of lift \cite{ref:nair2021}.  Let us examine the contours of $\nabla \times \bd{Z}$ superposed on $\omega$ for NACA0012 airfoil at $\alpha = 35^\circ$ for $\theta = 0$, $\pi/2$, $\pi$, $3\pi/2$ shown in figure~\ref{fig:Zwvort_airfoil_lift}.  We also consider the variation in lift coefficient with phase (shown in figure~\ref{fig:airfoil_lift}) to correlate phase advancement and delay with lift enhancement or reduction.  We observe positive $\nabla \times \bd{Z}$ very close to the leading edge and negative $\nabla \times \bd{Z}$ near the trailing edge of the airfoil at $\theta = 0$. This indicates that an injection of positive vorticity at the leading edge or an injection of negative vorticity at the trailing edge results in phase advancement, this might result in transient enhancement of lift. For the phases $\theta=\pi/2,\,3\pi/2$, which correspond to the extrema of $C_L$, we observe same sign of $\nabla \times \bd{Z}$ at both the leading and trailing edges; positive for $\pi/2$ and negative for $3\pi/2$.  At the phase of $\theta = \pi$ which corresponds to ${\rm avg}\,C_L$, we observe a negative $\nabla \times \bd{Z}$ near the leading edge and positive $\nabla \times \bd{Z}$ at the trailing edge. This is in contrast to the behavior observed at $\theta = 0$ as phase advancement now might result in a transient reduction in $C_L$ and a phase delay might result in transient lift enhancement. The injection of positive vorticity at the leading edge can lead to phase advancement at $\theta=0$  and phase delay at $\theta=\pi$, both through the modification of vortex formation dynamics can result in enhancement of lift. Such transient control based on the phase sensitivity function is also shown to increase the $C_L$ by Nair \textit{et al.} \cite{ref:nair2021}. Since vorticity injection into the flow is more challenging, we consider the associated $Z_u$ and $Z_v$ fields, which indicate that at, suction in $u$ direction near the leading edge might result in transient lift enhancement from the mean value ($\theta = 0,\,\pi$).  Hence, with the application of adjoint-based high fidelity phase sensitivity fields, we can correlate phase advancement or delay with the modification of vortex shedding patter over the airfoils, which further results in lift enhancement or reduction. The associated $Z_u$ and $Z_v$ fields can be used to select the location and direction for the flow control actuator to modify the vortex shedding behavior. 



\begin{figure}
    \centering
    \includegraphics[width=0.9\textwidth]{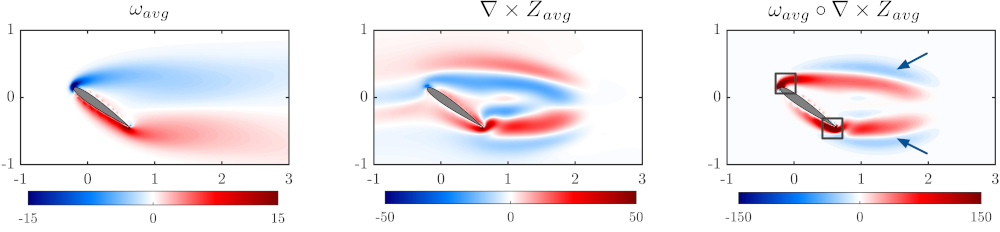}
    \caption{The phase averaged fields of vorticity ($\omega$), the phase-sensitivity function in terms of vector potential ($\nabla \times \bd{Z}$), and the Hadamard product ($\omega \circ \nabla \times \bd{Z}$) for flow over a NACA0012 airfoil at angle of attack $35^\circ$ for $Re = 100$.}
    \label{fig:airfoil_phaseavg}
\end{figure}

To further study the phase dynamic property, we show the phase-averaged vorticity and phase sensitivity fields in figure~\ref{fig:airfoil_phaseavg}.  We can observe that the averaged phase sensitivity field $\nabla \times \bd{Z}$ has highly sensitive regions near the trailing edge of the airfoil.  In order to assess whether these regions align with the vorticity field, we take the correlation (Hadamard product) of $\omega$ and $\nabla \times \bd{Z}$ are shown in figure~\ref{fig:airfoil_phaseavg} (right).  This product $\omega \circ \nabla \times \bd{Z}$ shows streak-like behavior near the regions of separation over the airfoil and the wake region with high shear, as highlighted in figure~\ref{fig:airfoil_phaseavg} (right).  This suggests that the vortex shedding can be influenced best near the leading and trailing edges of the airfoil as well as along the wake regions with high shear.

\section{Concluding remarks} \label{sec:4}

We have formulated an adjoint-based phase reduction analysis for periodic flows governed by the incompressible Navier--Stokes equations. We have derived the evolution equation for the phase sensitivity function that could be efficiently solved with any well-established numerical scheme through a single pair of forward and adjoint simulations in contrast to the direct method which requires numerous simulations. The properties of the spatial phase sensitivity function and their relations with the perturbations in the velocity potential, velocity, and vorticity fields are established. This enabled us to interpret the phase sensitivity fields for phase advancement or delay in terms of the velocity and vortical perturbations that are introduced to the flowfield. We have implemented this approach to perform phase sensitivity analysis for wakes of circular cylinder and symmetric airfoils of various thicknesses at post-stall angles of attack.

We have shown that this adjoint-based phase reduction method can be incorporated consistently into the immersed boundary projection method which can simulate flows over bodies of arbitrary shapes.  Using the immerse boundary projection method, we simulated two-dimensional laminar incompressible flow over a circular cylinder at $Re=100$ and validated the phase sensitivity functions obtained using the present framework with those obtained by the direct method~\cite{ref:khodkar20}. The phase for the cylinder flow is defined using the limit cycles oscillations of the lift coefficient i.e., $C_L$-$\dot{C}_L$ plane. The separation points on the cylinder are highlighted as the sensitive regions to perturbations consistent with the flow physics.  

Subsequently, we investigated the phase dynamic characteristics for wakes of symmetric NACA airfoils.  Chord-based thicknesses of $6\%,\,12\%$ and $20\%$ at angles of attack of $\alpha = 35^\circ,\, 45^\circ,\,55^\circ$ were considered at $Re=100$ to study the effect of angle of attack and thickness on phase sensitivity fields. These flows are characterized by their periodic vortex shedding behavior and the phase has been defined in the $C_L-\dot{C}_L$ plane. For all these cases, the leading and trailing edges of the airfoils have been highlighted as the sensitive regions for phase modification as these regions are the main sources of vorticity generation. The analysis revealed that an airfoil at a lower angle of attack is more sensitive to perturbations that cause phase advancement or delay than that at a higher angle of attack with the presence of large vortical structures in its wake. Furthermore, we observe that a thick airfoil has high magnitudes of phase sensitivity function making it sensitive to perturbations than a thin airfoil with a higher curvature.

Since the phase is defined based on the lift coefficient which closely tied to vortex dynamics, phase modification is achieved through the acceleration or deceleration of the shedding process. The phase advancement or delay leads to the lift enhancement or reduction caused by the modification of vortex formation dynamics over these bodies. Such open-loop lift-enhancement strategies are proposed based on the phase sensitivity function with respect to vorticity. For phases corresponding to mean lift coefficient, positive and negative vorticity injection at the leading and trailing edges of the airfoil might the enhancement of lift through phase advancement or delay at the particular phases. The phase sensitivity fields obtained through the adjoint-based formulation with reduced computational effort pave way for the development of reduced order closed loop control strategies for lift enhancement and drag reduction applications in periodic fluid flows and those with strong tonal dynamics.

\section*{Acknowledgments}
  Y.K. acknowledges financial support from JSPS (Japan) KAKENHI Grant Numbers JP20K03797, JP18H03205 and JP17H03279.
  Y.K. also acknowledges support from Earth Simulator JAMSTEC Proposed Project.
  K.T. acknowledges support from 
  the US National Science Foundation
  (Grant: 2129639)
  and the US Air Force Office of Scientific Research 
  (Grant: FA9550-21-1-0178).

\bibliographystyle{unsrt}  


\end{document}